\pdfoutput=1  
\documentclass[]{spie}  %

\usepackage{amsmath,amsfonts,amssymb}
\usepackage{graphicx}
\graphicspath{ {./graphics/} }
\usepackage{booktabs}  %
\usepackage{adjustbox}  %
\usepackage[table]{xcolor}
\usepackage[colorlinks=true, allcolors=blue]{hyperref}

\title{Evaluating the label efficiency of contrastive self-supervised learning for multi-resolution satellite imagery}

\author[a,b]{Jules Bourcier}
\author[b]{Gohar Dashyan}
\author[a]{Jocelyn Chanussot}
\author[a]{Karteek Alahari}
\affil[a]{Univ.\ Grenoble Alpes, Inria, CNRS, Grenoble INP, LJK, 38000 Grenoble, France}
\affil[b]{Preligens (ex-Earthcube), 75000 Paris, France}

\authorinfo{Further author information: (Send correspondence to J. Bourcier)\\J. Bourcier: E-mail: jules.bourcier@preligens.com}

\begin{document} 
\maketitle

\begin{abstract}

The application of deep neural networks to remote sensing imagery is often constrained by the lack of ground-truth annotations. Adressing this issue requires models that generalize efficiently from limited amounts of labeled data, allowing us to tackle a wider range of Earth observation tasks.
Another challenge in this domain is developing algorithms that operate at variable spatial resolutions, e.g., for the problem of classifying land use at different scales. Recently, self-supervised learning has been applied in the remote sensing domain to exploit readily-available unlabeled data, and was shown to reduce or even close the gap with supervised learning. In this paper, we study self-supervised visual representation learning through the lens of label efficiency, for the task of land use classification on multi-resolution/multi-scale satellite images. We benchmark two contrastive self-supervised methods adapted from Momentum Contrast (MoCo) and provide evidence that these methods can be perform effectively given little downstream supervision, where randomly initialized networks fail to generalize. Moreover, they outperform out-of-domain pretraining alternatives. We use the large-scale fMoW dataset to pretrain and evaluate the networks, and validate our observations with transfer to the RESISC45 dataset.

\end{abstract}


\section{Introduction}
\label{sec:intro}

The application of deep learning techniques to remote sensing imagery presents many challenges, one of the most important being the scarcity of annotations. Although large amounts of satellite imagery are readily available, curating and annotating them for a specific Earth observation tasks is usually very expensive, time-consuming and requires fine domain expertise. This implies that it is impractical in many real-world contexts to acquire the data needed to effectively leverage classical supervised learning methods. From this perspective, it is necessary to develop label-efficient approaches, i.e., models that are able to learn with few annotated samples.
Self-supervised learning (SSL) is a promising approach for this purpose, as it pretrains representations without requiring human labeling.
Inspired by the success of recent methods on natural images benchmarks~\cite{he2020momentum, chen2020simple, caron2020unsupervised, grill2020bootstrap, zbontar2021barlow}, SSL has been applied in the remote sensing domain to exploit the plentiful unlabeled data, and was shown to reduce or even close the gap with supervised learning and transfer from ImageNet~\cite{ayush2021geography, manas2021seasonal} (IN).

Another common problem in remote sensing is to process images covering various spatial scales,  e.g., for the task of classifying land use, where categories can range from individual storage tanks to full harbors. The capacity of SSL methods to generalize from few labels on this important problem was not explored by previous works, to the best of our knowledge.

To address this, in this paper we study self-supervised visual representation learning through the lens of label efficiency, for the classification of land use at different spatial resolutions. We benchmark two contrastive self-supervised methods adapted from Momentum Contrast (MoCo) \cite{he2020momentum} to assess their capacity to learn generic, multi-resolution features, which do not need many labeled examples for downstream image classification. We use the large-scale and diverse fMoW dataset \cite{christie2018functional} to pretrain and evaluate the networks, and validate our observations with transfer to the RESISC45 dataset \cite{cheng2017remote}, using diverse evaluation methods and amounts of labels.
We provide evidence that these methods can be trained effectively in few-label settings that are insufficient for randomly initialized networks to generalize. Thanks to MoCo with temporal positives, \cite{ayush2021geography}, when finetuning the pretrained models to RESISC45 with only about 4 examples per class, we reach 5× the accuracy of a classifier trained from scratch. Moreover, with simple linear probing on frozen representations, we surpass from-scratch networks in every label setting on fMoW and RESISC45. Additionally, the MoCo variants applied on fMoW images outperform out-of-domain pretraining on IN by significant margins, despite being pretrained on 3× less data. We also reveal that a basic $k$-nearest neighbors ($k$-NN) classifier on the learned representations provides out-of-the-box efficient generalization, and competes or outperforms finetuning with other methods when only few labels are available.

Our main contributions are as follows:
\vspace{-\topsep}
\begin{itemize}
    \item We experiment with two contrastive SSL methods, MoCo~\cite{he2020momentum} and MoCoTP~\cite{ayush2021geography}, on multi-resolution land use classification on the fMoW dataset, and observe their efficiency in terms of annotations required, on the common evaluation settings for representation learning with $k$-NN, linear, and finetuned classifiers.
    \item To demonstrate the transferability of the pretrained representations for land use prediction at different spatial resolutions, we study how the models pretrained on fMoW generalize to the smaller RESISC45 dataset, including extremely few images per class.
\end{itemize}

\section{Related work}
\label{sec:related-work}

\subsection{Self-supervised learning and contrastive learning}

SSL methods learn representations of data without relying on manual annotations. It consists in \textit{pretraining} a neural network to solve a \textit{pretext task} on unlabeled data, for the purpose of extracting semantic representations that allow for effective \textit{transfer} to downstream predictive tasks such as classification, segmentation or object detection. In computer vision, the popularity of SSL is due to recent methods that have shown to perform comparably well or even better than their supervised counterparts on natural image benchmarks \cite{he2020momentum, chen2020simple, caron2020unsupervised, grill2020bootstrap, zbontar2021barlow}.

\textit{Contrastive learning} has established itself as the staple framework for SSL of visual representations, with approaches such as MoCo \cite{he2020momentum}, SimCLR \cite{chen2020simple}, and SwAV \cite{caron2020unsupervised}. These methods work by attracting embeddings of pairs of sample images known to be semantically similar (\textit{positive pairs}) while simultaneously repelling pairs of dissimilar samples (\textit{negative pairs}). The most common way to define similarity is with the \textit{instance discrimination} pretext task \cite{wu2018unsupervised}, in which positives are generated as random data augmentations on the same image, and negatives are simply generated from different images. Thanks to this objective, the encoder learns to encode close representations for different views of the same object instance in an image and distant representations for other instances.
In this work, we use the strong contrastive method of MoCo~\cite{he2020momentum} as well as an extension proposed in Ref. \citenum{ayush2021geography} that adapts the learning objective to the spatio-temporal structure of satellite imagery.

\subsection{SSL in remote sensing}

Following their success in computer vision, several works have applied SSL methods to remotely sensed imagery for Earth observation tasks. Ref.~\citenum{jean2019tile2vec} was one of the first to use of contrastive learning for remote sensing representation learning. Ref.~\citenum{kang2021deep} applies a spatial augmentation criteria on top of MoCo~\cite{he2020momentum}. These works exploit a relevant assumption about the remote sensing domain: images that are geographically close should be semantically more similar than distant images.
Another way of making the learning procedure \textit{geography-aware} is to exploit the spatio-temporal nature of satellite imagery. Ref.~\citenum{ayush2021geography} uses spatially-aligned images over time to construct \textit{temporal positive pairs} with MoCo. The resulting temporally-aligned features were shown to improve generalization for classification, segmentation and object detection downstream tasks. In this work, we adopt this model with notable improvements, to study how such temporal invariance can benefit learning from few labels. In the same vein, Ref.~\citenum{manas2021seasonal} proposes a method that learns representations that are simultaneously variant and invariant to temporal changes.
One can also exploit the multi-spectral and multi-sensor nature of remote sensing. Ref.~\citenum{stojnic2021self} applies CMC~\cite{tian2020contrastive} on multi-spectral images, using different subsets of channels as augmented (positive) views.
Ref.~\citenum{swope2021representation} extends this to co-located images from multiple sensors, combining different sensor channels to construct positive pairs.

\subsection{Label efficiency of remote sensing representations}

Acquiring large amounts of labeled remote sensing data is often hard and prohibitively expensive. Surprisingly, few works have specifically studied how models are able to learn to solve Earth observation tasks with few annotated examples. The SSL method of Ref.~\citenum{manas2021seasonal} was shown to be useful for \textit{label-efficient} transfer learning on multi-label land cover classification on the BigEarthNet dataset~\cite{sumbul2019bigearthnet}. It was benchmarked on Sentinel-2 image chips from a fixed resolution of 10m and a rather different semantic domain (mainly land cover in natural environments) from the functional land use classification task of fMoW~\cite{christie2018functional} studied in the present work.
Ref.~\citenum{neumann2020indomain} shows that in-domain supervised pretraining can improve the performance in low-labeled settings on downstream classification datasets. However, it shows that transfer performance is very dependent on the labeling and data curation quality in the upstream dataset and leaves unresolved the problem of obtaining generic representations with less dependence on labels.
Ref.~\citenum{uzkent2019learning} proposes a weakly-supervised multi-modal pretraining method using paired satellite images and geo-located Wikipedia articles, and outperforms other pretraining strategies on fMoW classification when finetuned on small amounts of labeled data. However this is a rather different direction than SSL for learning representations and requires each satellite image to be paired to its corresponding article, which severely limits the amount of images that can be used.
In this work, we investigate contrastive SSL for label-efficient learning on multi-resolution imagery, which, to the best of our knowledge, has not been previously done. The three-faceted evaluation we perform with linear, finetuning and $k$-NN classification, is also more complete that what previous works have employed in their protocols.

\section{Method}
\label{sec:method}

\subsection{Datasets}

We study the task of land use classification from satellite images, on the fMoW \cite{christie2018functional} and RESISC45 \cite{cheng2017remote} datasets.

fMoW \cite{christie2018functional} is a large-scale dataset containing 363,571 RGB training images, across 62 fine-grained and diverse categories of functional land use. It provides several images from same locations over time, over 83,412 locations across 207 countries for training. Notably, the objects represented cover a varying range of ground resolution, from 0.5m for small structures (e.g., wind turbine) to 35m for large facilities (e.g., airport). As fMoW is a big, diverse, and multi-resolution dataset, we use it for self-supervised pretraining with the hope to learn rich semantic representations for remote sensing. We also use it for evaluation of the pretrained networks on the land use classification task with the included labels.

For complementarity, we selected RESISC45 for evaluation by transfer from fMoW pretraining to a separate downstream dataset. RESISC45 is of much smaller scale, with 18,900 RGB training images divided into 45 land use classes. The images are extracted from Google Earth from over 100 countries.
The categories are similar or overlap with the ones in fMoW, and the data is characterized by a multi-resolution distribution close to fMoW, ranging from 0.2 to 30m. Despite this, there is a degree of domain gap between fMoW and RESISC45, at least through some unrelated categories, which is interesting for the evaluation of the transferability of features pretrained on fMoW.

Therefore, these two datasets are well suited for studying label-efficient representation learning on the task of multi-resolution and multi-scale land use classification.

\noindent
\subsection{Models}

We employ the MoCo~\cite{he2020momentum} framework for contrastive SSL, specifically two different variants of the proposed MoCo-V2~\cite{chen2020mocov2} model. They differ in the selection of the positive views: we take (i) the usual MoCo setup of selecting two artificially augmented samples of the same image as positive views, and (ii) the MoCo with Temporal Positives (MoCoTP) setup~\cite{ayush2021geography}, which leverages the temporal resolution of geospatial data for generating positive pairs, i.e., positive views are sampled in temporal sequences of spatially aligned images, and the same augmentations of the usual MoCo setup are then applied to these temporal views. Compared to the MoCoTP framework from Ref.~\citenum{ayush2021geography}, we further add two improvements: (i) additional augmentations for rotational invariance; (ii) a fixed loss function that removes the false temporal negatives in the learning process. These improvements are detailed in Appendix~\ref{app:baseline-imp}.

We compare the two self-supervised methods against the following baselines: supervised pretraining on IN, pretraining with MoCo on IN, and random initialization (i.e., no pretraining). For RESISC45, we also add the baseline of supervised pretraining on fMoW (with the available land use labels), in addition to our SSL models on fMoW. We use a ResNet-50~\cite{he2016deep} encoder in all experiments.

The methods are evaluated for classification with three common evaluation procedures: (i) \textit{linear probing}, that trains a logistic regression on top of frozen features from the pretrained encoder that maps to the output logits for each class of the target task; (ii) \textit{finetuning}, which trains a logistic regression and also updates all the parameters of the network end-to-end with the target labels; (iii) \textit{$k$-NN}, that matches the frozen feature of an image to the $k$ nearest stored training features that votes for the label.

The SSL models are trained on all the images of the fMoW dataset (without labels). We then evaluate their performance against supervised baselines on the target classification task, by training a classifier on three subsets containing 1, 10, and 100\% of the labeled training data.
When sampling a subset of the full training set (1\% and 10\% labels), we conduct 3 Monte Carlo experiments, so that the results account for variance induced by data selection. Sampling is stratified, i.e., we ensure that the overall distribution of classes is maintained in every subset.

The implementation details for datasets and trainings are provided in Appendix~\ref{app:impl-details}.

\section{Results}
\label{sec:results}

\subsection{fMoW classification}
\label{sec:results:fmow-clf}

\subsubsection{Baseline improvements}
\label{sec:results:fmow-clf:baseline-imp}

We first show the performance enhancements brought by our modifications to the SSL framework of Ref.~\citenum{ayush2021geography}.
Table~\ref{tab:results-fmow-clf} shows the results of linear probing and finetuning on the 62-class land use classification task of fMoW of the methods with the proposed improvements, compared to the results from Ref.~\citenum{ayush2021geography}. Models are evaluated with single-image top-1 F1-score averaged over classes as well as accuracy on the validation set of fMoW.

With 100\% labels, we see that our improved reproduction of MoCoTP increases performance by 4.36\% F1 compared to Ref.~\citenum{ayush2021geography} in linear probing and 1.62\% in finetuning. This shows that the use of the rotation augmentations and the correction of false negatives in the loss function are helpful, especially for linear probing, which closes the gap with finetuning completely. As a note, the additional augmentations also improve all other baselines, e.g. it improves random initialization by 1.29\% F1, showing the general relevance of rotational invariance for satellite imagery.

\begin{table}[tb]
    \centering
    \begin{adjustbox}{max width=\textwidth}  %
    \begin{tabular}{l|cc|cc}
    \toprule
    ~ & \multicolumn{2}{c|}{F1-score} & \multicolumn{2}{c}{Accuracy} \\
    \cmidrule{2-5}
    ~ & Linear & Finetune & Linear & Finetune \\
    Method & $(\dag~/~\ast)$ & $(\dag~/~\ast)$ & $(\dag~/~\ast)$ & $(\dag~/~\ast)$ \\
    \midrule
    Random init & - &  64.71 / 65.39 & - &  69.05 / 69.33 \\
    IN-sup  &  ~~~-~~~ / 50.25 &  64.72 / 66.01 &  ~~~-~~~ / 54.56 &  69.07 / 70.80 \\
    IN-MoCo  &  31.55 / 53.47 &  57.36 / 65.37 &  37.05 / 57.28 &  62.90 / 70.17 \\
    MoCo        &  55.47 / 65.55  &  60.61 / 67.23 &  60.69 / 69.62 &  64.34 / 71.82 \\
    MoCoTP      &  \textbf{64.53} / \textbf{68.89} &  \textbf{67.34} / \textbf{68.96} &  \textbf{68.32} / \textbf{72.56} &  \textbf{71.55} / \textbf{73.01} \\
    \bottomrule
    \end{tabular}
    \end{adjustbox}
    \caption{Baseline improvements on fMoW classification. We report top-1 F1-score and accuracy (in \%) on the fMoW validation set, on linear probing and finetuning. Methods are different pretrainings: ``IN-" means using ImageNet transfer, ``sup" means supervised pretraining, and ``Random init" represents no pretraining. ``$\dag$" denote results from Ref.~\citenum{ayush2021geography}. ``$\ast$" denote our improved reproductions.}
    \label{tab:results-fmow-clf}
\end{table}

\subsubsection{Label efficiency}
\label{sec:results:fmow-clf:labeleff}

We now study label-efficient classification on fMoW by evaluating the models with $k$-NN, linear, and finetuning classifiers, for fractions of 1, 10, and 100\% labeled training data.
Table~\ref{tab:results-fmow-clf-all} shows the results. Figure~\ref{fig:results-fmow-clf-f1} in the appendix provides a graphical view of the same results.

On linear probing, we observe that, while the supervised baseline shows poor generalization from few labels, the MoCo variants retain a much higher performance in comparison as the fraction of labels is reduced. On the fewest number of labels, MoCo and MoCoTP respectively give +30 and +35\% accuracy against random initialization. In the semi-supervised settings of 1\% and 10\% labels, we see that MoCoTP respectively give 96\% and 87\% of the performance reached by 100\% labels, and furthermore outperforms the IN-pretrained methods by large margins.
These results indicate that in-domain contrastive SSL with the use of temporal priors is very efficient at learning semantic features from the upstream dataset.

When finetuning, interestingly, we see that MoCo and MoCoTP do not significantly improve performance against linear probing on the 1\% and 10\% settings. This means that they have learned to the best of their capacity to represent the classes linearly, which implies high-level semantic representations. On the contrary, IN-pretrained models are significantly improved by finetuning due to the domain gap between IN and fMoW; even so, they show inferior label efficiency that MoCo and MoCoTP, e.g., IN-supervised weights retains only 60\% of its maximum performance on 1\% of labels vs.\ 87\% for MoCoTP.
Perhaps surprisingly, we also remark that both MoCo and MoCoTP improve over the all supervised counterparts in the many-label regime of 100\%, e.g., by +1.84\% and +3.57\% F1 respectively vs.\ random initialization, showing than SSL can also be useful if we have large quantities of annotations for a task.

Finally, under the $k$-NN evaluation, we see that MoCoTP outperforms all other methods, and that for 1\% and 10\% labels, it provides even better classifier than other pretrainings with finetuned networks -- e.g., on 1\% fMoW, MoCoTP with $k$-NN is better than finetuned IN-supervised weights by 18.82\% F1. This further confirms that MoCoTP produces high quality and label-efficient features.

\begin{table}[tb]
    \centering
    \begin{adjustbox}{max width=\textwidth}  %
    \begin{tabular}{l|ccc|ccc|ccc}
    \toprule
    {} & \multicolumn{3}{c|}{1\% labels} & \multicolumn{3}{c|}{10\% labels} & \multicolumn{3}{c}{100\% labels} \\
    Method &           $k$-NN &        Linear &      Finetune &           $k$-NN &        Linear &      Finetune &    $k$-NN & Linear & Finetune \\
    \midrule
    Random init &             - &             - &  19.29 (1.65) &             - &             - &  51.87 (0.50) &      - &      - &    65.39 \\
    \hline
    IN-sup       &  21.04 (0.48) &  32.41 (0.17) &  39.43 (1.53) &  31.92 (0.25) &  43.86 (0.07) &  57.32 (0.07) &  39.65 &  50.25 &    66.01 \\
    IN-MoCo      &  22.59 (0.19) &  35.10 (0.26) &  37.93 (0.75) &  32.71 (0.40) &  46.85 (0.47) &  56.73 (0.52) &  40.72 &  53.47 &    65.37 \\
    \hline
    MoCo         &  47.34 (0.22) &  53.61 (0.30) &  52.38 (0.45) &  52.89 (0.15) &  61.13 (0.18) &  60.80 (0.31) &  57.57 &  65.55 &    67.23 \\
    MoCoTP       &  \textbf{56.75 (0.41)} &  \textbf{60.05 (0.11)} &  \textbf{60.00 (0.43)} &  \textbf{61.58 (0.09)} &  \textbf{66.15 (0.11)} &  \textbf{66.35 (0.75)} &  \textbf{64.86} &  \textbf{68.89} &    \textbf{68.96} \\
    \bottomrule
    \end{tabular}
    \end{adjustbox}
    \caption{Label-efficient land use classification on fMoW. We report top-1 F1-score (in \%) on the fMoW validation set, for 1, 10 and 100\% labeled training data, with $k$-NN, linear probing and finetuning classifiers. Methods are different pretrainings: ``IN-" means using ImageNet transfer, ``sup" means supervised pretraining, and ``Random init" represents no pretraining. For 1\% and 10\% labels, the values are `mean (sd)' of 3 runs with random sampling of a subset of the full training set.}
    \label{tab:results-fmow-clf-all}
\end{table}

\subsection{Transfer to RESISC45 classification}
\label{sec:results:resisc-clf}

A main goal of representation learning is to improve generalization on new tasks and datasets through feature re-use. We then evaluate label-efficient transfer learning on the RESISC45 dataset, with the same three classifiers and fraction of labeled samples as in Sec.~\ref{sec:results:fmow-clf:labeleff}. 
Table~\ref{tab:results-resisc45-clf-all} shows the results on the 45-class land use classification task of RESISC45. Models are evaluated with top-1 accuracy on the testing set. Figure~\ref{fig:results-resisc45-clf-acc} in the appendix provides a graphical view of the same results.

In the case of linear probing, we see that MoCo and MoCoTP pretrained on fMoW are label efficient with features adapted for linear classification on this different downstream dataset. While the supervised end-to-end baseline shows a large gap of performance of 74.83\% accuracy between training of 100 and 1\% of labels, due to the relatively small size of the dataset, the accuracy of MoCoTP for 1\% of labels is only  23.64\% accuracy lower than for 100\% of labels, achieving 68.93\%.

When finetuning, self-supervised features do not provide a significant advantage against IN pretraining when increasing the percentage of labeled data: we see that all the pretrained networks saturate on the full dataset at around 95\% accuracy. Nevertheless, fMoW-MoCoTP pretraining is more label-efficient than other methods in the very-low label regime of 1\%, outperforming IN-supervised pretraining by 5.28\% accuracy. This shows the capability of specialized pretraining with MoCoTP to improve generalization from very few labels, as in this setting there is only 175 examples, about 4 examples per class.

With $k$-NN classifiers, fMoW-MoCoTP features provide the best performance over all annotations regimes. When given 10\% of labels, it is on par with fMoW-supervised pretraining on 100\% labels, with 83.16 vs.\ 83.65\% accuracy. Furthermore, on 100\% of labels, it gives similar performance to from-scratch training (88.37 vs.\ 88.59\% accuracy). Note that $k$-NN is the simplest and weakest classifier one can build on top of pretrained networks.
These results indicate that in-domain SSL with temporal priors is effective and label-efficient for out-of-the-box transfer to a new downstream dataset.

\begin{table}[tb]
    \centering
    \begin{adjustbox}{max width=\textwidth}  %
    \begin{tabular}{l|ccc|ccc|ccc}
    \toprule
    {} & \multicolumn{3}{c|}{1\% labels} & \multicolumn{3}{c|}{10\% labels} & \multicolumn{3}{c}{100\% labels} \\
    Method &           $k$-NN &        Linear &      Finetune &           $k$-NN &        Linear &      Finetune &    $k$-NN & Linear & Finetune \\
    \midrule
    Random init &             - &             - &  13.76 (3.83) &             - &             - &  53.07 (1.04) &      - &      - &    88.59 \\
    \hline
    IN-sup       &  49.45 (1.20) &  53.91 (0.20) &  64.87 (1.27) &  70.51 (0.27) &  77.01 (0.34) &  87.44 (0.33) &  79.32 &  86.48 &    95.38 \\
    IN-MoCo      &  43.67 (1.28) &  49.79 (0.50) &  59.45 (1.04) &  65.94 (0.32) &  77.21 (0.53) &  84.98 (0.84) &  75.40 &  87.32 &    \textbf{95.41} \\
    \hline
    fMoW-sup     &  62.12 (2.11) &  65.86 (1.70) &  67.66 (2.17) &  77.13 (0.37) &  81.42 (0.34) &  88.14 (0.63) &  83.65 &  87.94 &    94.14 \\
    fMoW-MoCo    &  60.23 (0.60) &  64.22 (1.20) &  65.94 (1.00) &  79.61 (0.46) &  84.33 (0.32) &  87.69 (0.33) &  85.68 &  91.40 &    94.84 \\
    fMoW-MoCoTP  &  \textbf{65.47 (1.10)} &  \textbf{68.93 (0.86)} &  \textbf{71.15 (0.33)} &  \textbf{83.16 (0.17)} &  \textbf{87.18 (0.28)} &  \textbf{89.16 (0.15)} &  \textbf{88.37} &  \textbf{92.57} &    95.30 \\
    \bottomrule
    \end{tabular}
    \end{adjustbox}
    \caption{Label-efficient land use classification on RESISC45. We report top-1 accuracy (in \%) on the RESISC45 testing set, for 1, 10 and 100\% labeled training data, with $k$-NN, linear probing and finetuning classifiers. Methods are different pretrainings: ``IN-" means transfer from ImageNet, ``fMoW-" means transfer from fMoW, ``sup" means supervised pretraining, and ``Random init" represents no pretraining. For 1\% and 10\% labels, the values are `mean (sd)' of 3 runs with random sampling of a subset of the full training set.}
    \label{tab:results-resisc45-clf-all}
\end{table}

\section{Conclusion}
\label{sec:conclu}
We present a study of the annotation-efficiency of self-supervised contrastive learning for remote sensing, on the specific problem of multi-resolution and multi-scale land use classification. The results demonstrate the potential of methods based on MoCo, which provides a generalization capability from few labels that is not achievable with classical supervised predictors. We show that the use of MoCo with temporal positives further improves label-efficient learning.
Our observations indicate that SSL is a promising direction for solving Earth observation tasks that have previously been inaccessible due to the scarcity of annotations.  %
We can suppose that improvements to label efficiency could otherwise be promoted by scaling up the amount of pretraining data, performing more pretraining epochs, or using higher-capacity network architectures.
Further work could include additional studies such as extending the sampling of sizes for the downstream datasets, investigating the multi-scale aspects of features, expanding to object detection tasks, and experimenting with different SSL methods, such as SwAV~\cite{caron2020unsupervised} or BYOL~\cite{grill2020bootstrap}.

\appendix    %

\section{Graphical view of label-efficient classification}
\label{sec:graphic-labeleff-clf}

\begin{figure}[hbt!]
    \centering
    \includegraphics[width=\columnwidth]{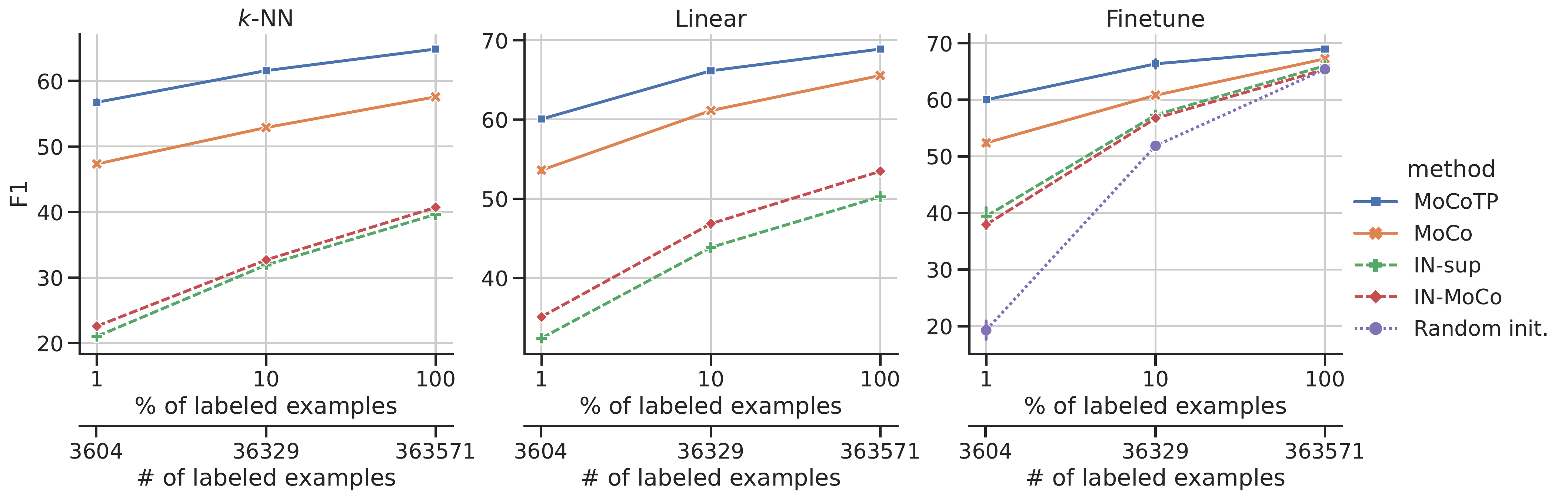}
    \caption{Label-efficient land use classification on fMoW. These values correspond to Tab.~\ref{tab:results-fmow-clf-all} in the main text, see the table caption for description.}
    \label{fig:results-fmow-clf-f1}
\end{figure}

\begin{figure}[hbt!]
    \centering
    \includegraphics[width=\columnwidth]{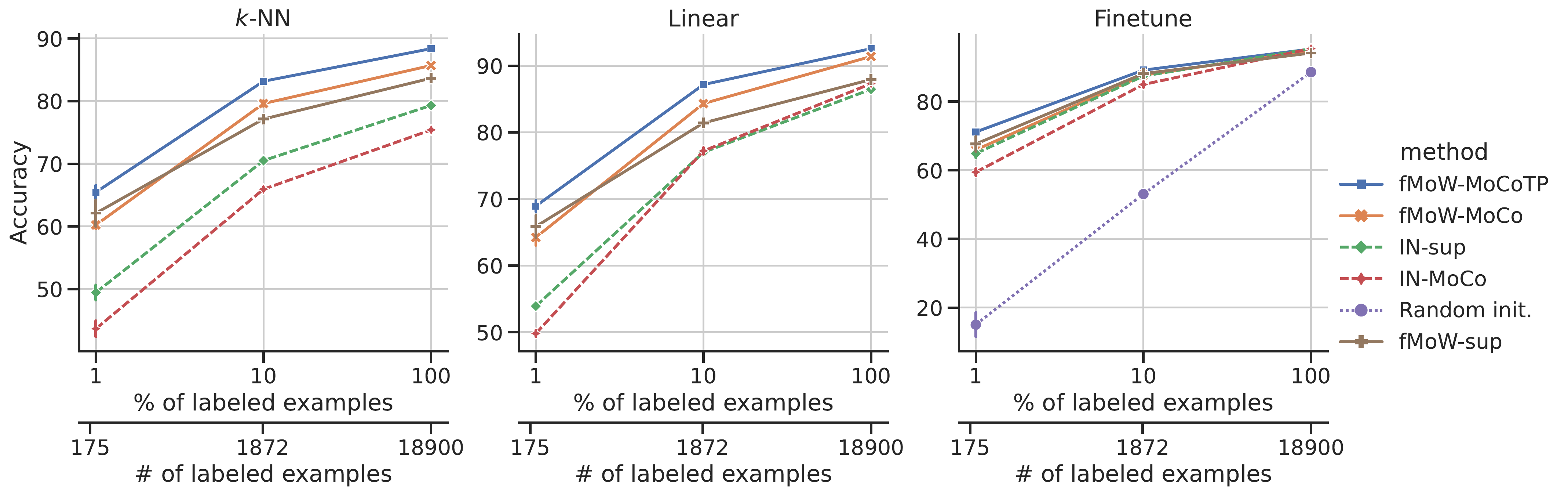}
    \caption{Label-efficient land use classification on RESISC45. These values correspond to Tab.~\ref{tab:results-resisc45-clf-all} in the main text, see the table caption for description.}
    \label{fig:results-resisc45-clf-acc}
\end{figure}

\section{Baseline improvement details}
\label{app:baseline-imp}

Compared to the framework of Ref.~\citenum{ayush2021geography}, we add two modifications to improve the models in our experiments.

\subsection{Additional augmentations}

In addition to the geometric and color perturbations of MoCo-V2, we apply random horizontal flips and rotations by multiples of 90°. Since the data augmentation scheme plays a leading role in contrastive learning \cite{tian2020makes}, we aim to learn representations more suited to overhead images thanks to rotational invariance. These additional augmentations are also used for supervised finetuning of all MoCo models and baselines. 

\subsection{False temporal negatives removal in MoCoTP}

MoCo learns to match an input \textit{query} $q$ to a \textit{key} $k^+$ (representing the encoded views of the same sample) among a set of negative keys ${k^-}$, using the instance discrimination pretext task \cite{wu2018unsupervised}--we refer readers to Ref.~\citenum{he2020momentum} for details on MoCo.
MoCo uses the popular choice of InfoNCE~\cite{oord2018representation} for the contrastive loss:
\begin{equation}\label{eq:infonce}
      \mathcal{L}(q, k^+) = -\log\frac
      {\exp{(q \cdot k^+/\tau)}}
      {\exp{(q \cdot k^+ / \tau)}
       + \sum_{k^-} \exp{(q \cdot k^- / \tau)}}
\end{equation}

\noindent
where $\tau$ is a temperature scaling parameter.

MoCoTP extends the instance discrimination pretext task to use spatially aligned images from different times as positives. This is implemented as a drop-in replacement for $q$ and $k^+$ in Eq.~\eqref{eq:infonce}. However, this can introduce \textit{false negatives}. Indeed, at each iteration of training, it may happen that the set of negatives ${k^-}$ contains temporal views for samples of the current mini-batch of queries. Such false negatives will cause an incorrect repulsion between the embeddings of similar samples. To what extent this is detrimental to the learned representations depends on the probability of sampling temporal pairs in the training set, as well as on the size of the queue. To avoid the false negatives to interfere with the learning objective, we simply mask out the logits $q \cdot k^-$ in the InfoNCE loss in \eqref{eq:infonce} for every $k^-$ that happens to be a temporal view of $q$.

\section{Implementation details}
\label{app:impl-details}

Here we describe the datasets and training settings used for experiments.

\subsection{Datasets details}

\subsubsection{fMoW}
Following Refs.~\citenum{christie2018functional} and \citenum{ayush2021geography}, we use the fMoW-RGB products in our experiments, which provides 3-bands imagery at 0.5m ground resolution. We use the official train and validation splits, composed of 363,571 images and 50,041 images respectively. Preprocessing is applied identically to Ref.~\citenum{christie2018functional} and input images are resized to 224×224 pixels.

\subsubsection{RESISC45}
We use the 224×224 images of RESISC45 without specific preprocessing. We use the train/validation/test splits defined in Ref.~\citenum{neumann2020indomain}, composed of 18,900, 6,300 and 6,300 images respectively.

\subsection{Training details}

In all our experiments, we use a ResNet-50 architecture for the encoder.
For linear, finetuning and $k$-NN evaluation, all our hyperparameters apply to both fMoW and RESISC45 experiments, and are the same regardless of the training subset size.
We use the \texttt{PyTorch} framework in our code which is based on the official implementation of Ref.~\citenum{ayush2021geography}\footnote{\url{https://github.com/sustainlab-group/geography-aware-ssl}}. All trainings are performed on compute nodes with 4 Tesla V100 GPUs.

\subsubsection{MoCo pretraining}
Pretraining with the two MoCo variants is performed with the following hyperparameters: learning rate of 3e-2 with a cosine schedule, batch size of 256, dictionary queue size of 65536, temperature scaling of 0.2, SGD optimizer with a momentum of 0.9, weight decay of 1e-4. We use MoCo-V2 as the contrastive framework~\cite{chen2020mocov2}. The data augmentation scheme is the random composition of resized cropping, horizontal flipping, color jittering, Gaussian blur, horizontal flipping, and 90° rotations. Pretraining is performed for 200 epochs.

\subsubsection{Linear probing}
For linear evaluations, we train a supervised linear classifier on top of frozen features from the output of the ResNet global average pooling layer.
For MoCo methods, we use a learning rate of 1 reduced by a multiplicative factor of 0.5 when the validation loss plateaus for 5 epochs, a batch size of 256, no weight decay, and only random resized cropping for the augmentations.
For supervised methods, hyperparameters are identical except we use a learning rate of 1e-3.
All models are trained with cross-entropy loss until convergence of the validation loss, and evaluated on epoch with the best top-1 accuracy on the validation set.

\subsubsection{Finetuning}
For finetuning evaluations, we initialize networks with the pretrained weights and adapt them during training.
For MoCo methods, we use a learning rate of 3e-4 for ResNet weights and 1 for the final classification layer, reduced reduced by a multiplicative factor of 0.5 when the validation loss plateaus for 2 epochs; a batch size of 256, no weight decay, and the same augmentations used for MoCo pretraining.
For supervised methods, hyperparameters are identical except we use a single learning rate of 1e-3 and weight decay of 1e-4.
All models are trained with cross-entropy loss until convergence of the validation loss, and evaluation on epoch with the best top-1 accuracy on the validation set.\noindent

\subsubsection{$k$-NN}
We adopt the weighted nearest neighbors classifier of Ref.~\citenum{wu2018unsupervised}, following common practice. We freeze the pretrain model to compute features at the output of the ResNet global average pooling layers. We do not apply any data augmentation. The classifier has only one hyperparameter: the number of nearest neighbors $k$. We tune $k$ and find that a value of 200 consistently works best for our MoCo methods; we then use this value across all experiments.

\subsubsection{ImageNet-pretrained weights}
We use available pretrained networks on IN-1K~\cite{deng2009imagenet} for baselines. For the IN-supervised method, we use the standard weights available in the \texttt{torchvision}\footnote{\url{https://pytorch.org/vision}} library.
For the IN-MoCo method, we take the official weights of MoCo-V2 pretrained for 200 epochs\footnote{available at \url{https://github.com/facebookresearch/moco}}.

\acknowledgments %

We thank Thomas Floquet, Tugdual Ceiller, and other colleagues at Preligens for fruitful discussions and for providing important feedback.
Karteek Alahari was supported in part by the ANR grant AVENUE (ANR-18-CE23-0011).
This work was granted access to the HPC resources of IDRIS under the allocation 2021-AD011013097 made by GENCI.

\bibliography{references} %
\bibliographystyle{spiebib} %

\end{document}